\documentclass[aps,prl,twocolumn,floats,floatfix]{revtex4-1}

\usepackage{graphicx}
\usepackage{amsmath,amssymb}
\usepackage{psfrag}
\usepackage{color}

\begin{document}

\newcommand{\gapp}{g_{\scriptscriptstyle \rm approx}}
\newcommand{\kB}{k_{\scriptscriptstyle \rm B}}
\newcommand{\TK}{T_{\scriptscriptstyle \rm K}}
\newcommand{\Tc}{T_{\scriptstyle \rm c}}
\newcommand{\Phic}{\Phi_{\scriptstyle \rm c}}
\newcommand{\fc}{f_{\scriptstyle \rm c}}

\newcommand{\tin}{t_{\scriptscriptstyle \rm in}}
\newcommand{\tout}{t_{\scriptscriptstyle \rm out}}

\newcommand{\tint}{t_{\scriptscriptstyle \rm int}}
\newcommand{\ts}{t_{\scriptstyle \sigma}}
\newcommand{\hs}{h_{\scriptstyle \sigma}}
\newcommand{\cs}{c_{\scriptstyle \sigma}}
\newcommand{\phiEA}{\phi_{\scriptscriptstyle \rm EA}}
\newcommand{\M}{\mathsf M}

\newcommand{\MCT}{{\small MCT}}
\newcommand{\CFS}{{\small CFS}}
\newcommand{\RFOT}{{\small RFOT}}

\newcommand{\beq}{\begin{equation}} \newcommand{\eeq}{\end{equation}}
\newcommand{\bea}{\begin{eqnarray}} \newcommand{\eea}{\end{eqnarray}}
\newcommand{\wh}{\widehat} \newcommand{\wt}{\widetilde}
\def\to{\rightarrow} \def\la{\left\langle} \def\ra{\right\rangle}

\let\a=\alpha \let\b=\beta \let\g=\gamma \let\d=\delta
\let\e=\varepsilon \let\z=\zeta \let\h=\eta \let\k=\kappa
\let\l=\lambda \let\m=\mu \let\n=\nu \let\x=\xi \let\p=\pi
\let\s=\sigma \let\t=\tau \let\f=\varphi \let\ph=\varphi\let\c=\chi
\let\ps=\psi \let\y=\upsilon \let\si=\varsigma \let\G=\Gamma
\let\D=\Delta \let\Th=\Theta\let\L=\Lambda \let\X=\Xi \let\P=\Pi
\let\Si=\Sigma \let\F=\Phi \let\Ps=\Psi \let\Y=\Upsilon
\let\ee=\epsilon \let\r=\rho \let\th=\theta \let\io=\infty
\let\om=\omega


\title{ Crossover from $\beta$ to $\alpha$ relaxation in cooperative
  facilitation dynamics}

\author{Mauro Sellitto} \affiliation{Dipartimento di Ingegneria
  Industriale e dell'Informazione, Seconda Universit\`a di Napoli,
  \\ Real Casa dell'Annunziata, Via Roma 29, I-81031 Aversa (CE),
  Italy}

\begin{abstract}
  $\beta$ and $\alpha$ relaxation processes are dynamical scaling
  regimes of glassy systems occurring on two separate time scales
  which both diverge as the glass state is approached.  We study here
  the crossover scaling from $\beta$- to $\alpha$- relaxation in the
  cooperative facilitation scenario (\CFS) and show that it is
  quantitatively described, with no adjustable parameter, by the
  leading order asymptotic formulas for scaling predicted by the
  mode-coupling theory (\MCT).  These results establish (i) the
  mutual universality of the \MCT\ and \CFS, and (ii) the existence
  of a purely dynamic realization of \MCT\ which is distinct from the
  well-established random first order transition scenario for
  disordered systems.  Some implications of the emerging
  kinetic-static duality are discussed.

\end{abstract}

\maketitle

The glass state is obtained with striking simplicity in a variety of
substances~\cite{BiKo}.  Yet, its fundamental nature is one of the
most enduring puzzles of condensed matter physics.  From a dynamical
point of view, what makes the glass relaxation so peculiar is the
existence of {\em two} time scales which both grow dramatically as the
separation $\sigma$ from the glass state vanishes.  The first scale,
$t_{\sigma}$, refers to the rattling motion of particles in the cage
formed by their neighbors ($\beta$ relaxation).  The second one,
$t_{\sigma}' \gg t_{\sigma}$, is generally ascribed to the slower
cooperative rearrangements of cages that allow long-term particle
diffusion ($\alpha$ relaxation).  This two-step relaxation pattern was
first predicted by mode-coupling theory (\MCT)~\cite{Goetze_book}, and
entails as precursors phenomena strongly temperature or density
dependent spectra exhibiting relaxation stretching.  Pioneer work
first identified these glass precursors in colloidal suspensions of
hard spheres~\cite{Pusey} and in molecular-dynamics simulations for a
Lennard-Jones mixture~\cite{KoAn}.

The crossover from $\beta$ to $\alpha$ relaxation, which occurs for
times large on scale $\ts$ but small on scale $\ts'$, constitutes a
crucial test for our understanding of glassy dynamics~\cite{Goetze90}.
It is remarkable because the ratio $t_{\sigma}'/t_{\sigma}$ is itself
diverging as $\sigma \to 0$.  At the same time, however, its actual
investigation is delicate because of the sensitive dependence on three
quantities: the separation parameter $\sigma$, the system-specific
exponent parameter $\lambda$, and the arrested part of correlation,
$f$.  In realistic systems these quantities can be inferred only with
limited accuracy (or may even be not well defined). This makes it
difficult to assess unambiguously the \MCT\ predictions and only a few
cases have been analyzed~\cite{Franosch,Gleim}.  In this Letter we
provide a test of the crossover scaling in the cooperative
facilitation scenario (\CFS)~\cite{FrAn}.  Several results have
already suggested a close analogy between \MCT\ and
\CFS~\cite{SeBiTo,ArSe,Se2012,FrSe}. Here we show that this relation
is, in fact, quantitative and deep.

{\bf \MCT\ scaling laws}. -- Without loss of generality we refer to
the ideal version of \MCT, in which the normalized density-fluctuation
correlator $\Phi(t)$ has no wave-vector dependence (for an exhaustive
description of \MCT, see Ref.~\cite{Goetze_book}).
Near the glass bifurcation singularity the correlator exhibits a
plateau at some $\fc$.  Near this plateau the small-$\sigma$ dynamics
on scale $\ts$ is ruled by the {\em first scaling law}:
\begin{eqnarray}
\Phi(t) - \fc = \cs \, g(\hat t), \qquad \hat t = \frac{t}{\ts}.
\label{eq_1st}
\end{eqnarray}
Here the correlation scale $\cs$ and the time scale $\ts$ read
\beq \cs \propto \sqrt{|\s|}, \qquad \ts \propto |\s|^{-\frac{1}{2a}},
\label{eq_cs}
\eeq
where $a$ is some critical exponent. The $\sigma$-independent master
function $g(\hat t)$ obeys the scale-invariant equation
\beq - 1 + \lambda \, g(\hat t)^2 - \partial_{\hat t} \int_0^{\hat t}
g(\hat t - \hat t') \, g(\hat t') \, \mathrm{d}\hat t' = 0
, \label{eq_g} \eeq
to be solved for the initial condition $\lim_{\hat t \to 0} g(\hat t)
\, \hat{t}^a$ = 1. The exponent parameter, $\lambda$, quantifies all
properties of $g(\hat t)$. In particular, it determines the exponent
$a$ via
\beq \frac{\Gamma(1-a)^2}{\Gamma(1-2a)} = \lambda.
\label{lambda_a}\eeq
%
%
Two time regions can be distinguished in the first scaling regime
where, depending on whether $\Phi$ is above or below $\fc$, the master
function $g(\hat t)$ can be approximated as
\bea g(\hat t) \simeq
\left\{
\begin{array}{ll} 
 g_a(\hat t) = \hat{t}^{-a} - A_1 \, \hat{t}^a,
 \,\,\,\,\,\,\,\,\,\,\,\,\,\, \mbox{for \,} \Phi > \fc \\ \\ g_b (\hat
 t) = - B \ \hat{t}^{b} + \frac{B_1}{B} \, \hat{t}^{-b}, \,\,\,\,
 \mbox{for \,} \Phi < \fc
\end{array}
\right.  \label{eq_gab} \eea with an error decreasing faster than
$\hat{t}^a$ in the former case and $\hat{t}^{-b}$ in the latter.  The
coefficients $A_1$ and $B_1$ read
\bea \frac{1}{2 A_1} &=& \Gamma(1-a) \, \Gamma(1+a)-\lambda
 \label{eq_A1} \\ \frac{1}{2 B_1} &=& \Gamma(1-b) \,
\Gamma(1+b)-\lambda, \label{eq_B1} \eea
while $B$ is a positive quantity fixed only by $\lambda$. The dominant
contribution to $g_b(\hat t)$, $-B \hat{t}^b$, is called von
Schweidler decay. The von Schweidler exponent $b$ is determined by
\beq \frac{\Gamma(1+b)^2}{\Gamma(1+2b)} = \lambda,
\label{lambda_b}\eeq
with $0< b \le 1$.  The function $g(\hat t)$ exhibits a zero at some
$\hat{t}^*$.  The part for $\hat t $ preceding $\hat{t}^*$ deals with
deviation from the critical decay. The part for $\hat t > \hat{t}^*$
describes the approach toward the von Schweidler decay.  The
quantities $\hat{t}^*$ and $B$ have to be calculated by solving
Eq.~(\ref{eq_g}). The numerical solution~\cite{Goetze90} shows that
there is an interval for $\hat t$ centered by $\hat{t}^*$, where the
three functions $g_a,\, g_b$ and $g$ are very close to each
other. Therefore, to locate approximately $\hat{t}^*$ and $B$ we set
$g_a(\hat{t}^*)=g_b(\hat{t}^*) = 0$, which gives $\hat{t}^* \simeq
A_1^{-1/2a}$ and $B \simeq \sqrt{A_1^{b/a} \, B_1}$.  These
approximations permit us to base the comparison with the \CFS\ on
elementary formulas.

\begin{figure}
\includegraphics{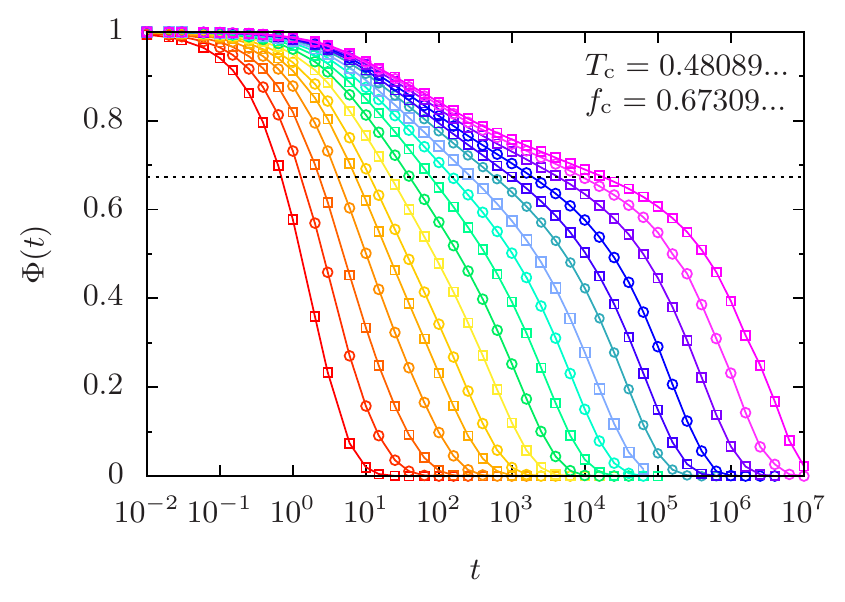}
\vspace{-.5cm}
\caption{Correlation decay for temperature $T \in [0.49,\,10]$ in the
  \CFS\ on a Bethe lattice with coordination $z=4$ and facilitation
  parameter ${\sf f}=2$.  System size $N=2^{23-24}$.}
\label{fig.phi_raw}
\end{figure}

\begin{figure}
\includegraphics{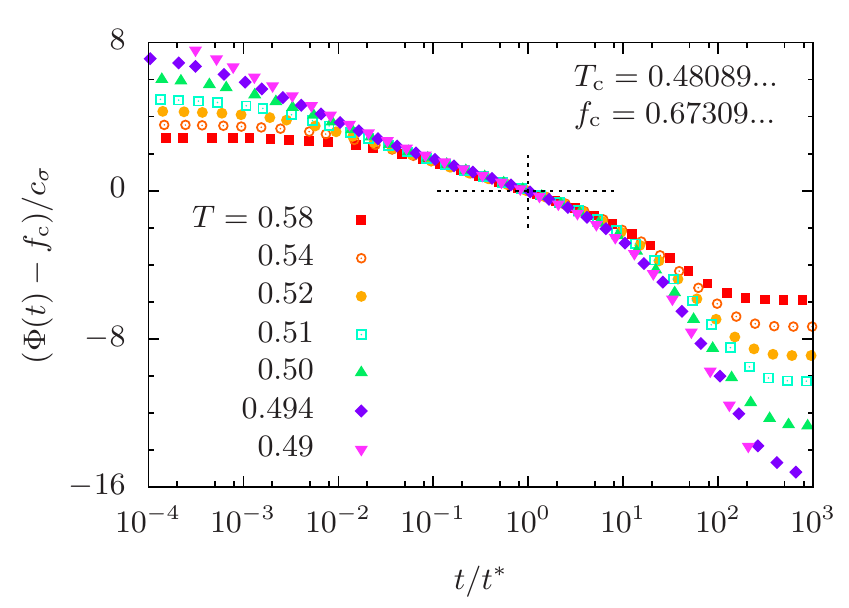}
\vspace{-.5cm}
\caption{\MCT\ first scaling law in the \CFS. Rescaled correlator vs
  rescaled time variable $\hat{t}/\hat{t^*}=t/t^*$.}
\label{fig.scaling1}
\end{figure}

\begin{figure}
\includegraphics{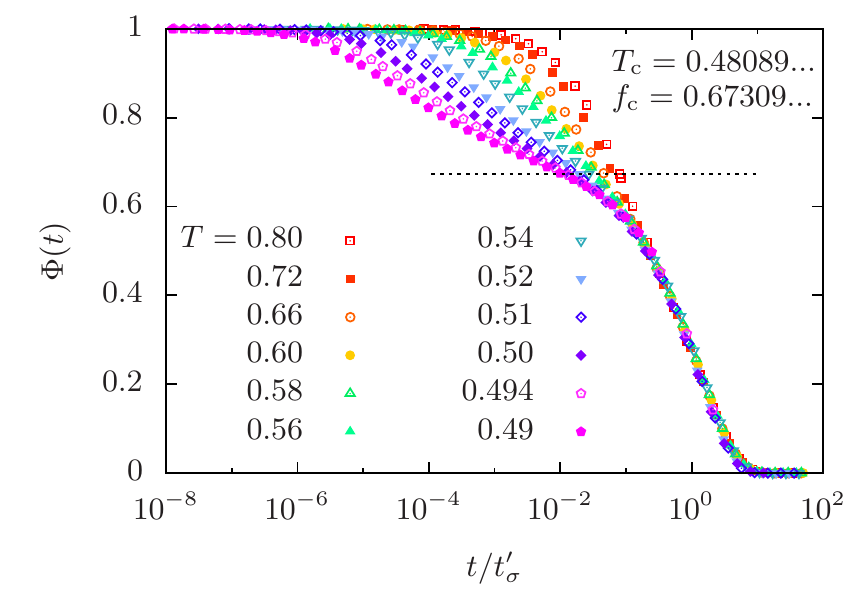}
\vspace{-.5cm}
\caption{\MCT\ second scaling law in the \CFS.}
\label{fig.scaling2}
\end{figure}

Substituting the von Schweidler decay into Eq.~(\ref{eq_1st}), one
gets the von Schweidler decay law
\beq \Phi(t) - \fc = - \left( \frac{t}{\ts'} \right)^b, \eeq
where the new relevant time scale is now
\bea \ts' \propto |\s|^{-\gamma}, \qquad \gamma=\frac{1}{2a} +
\frac{1}{2b} . \label{gamma_ab} \eea
von Schweidler's law describes the small-$\s$ dynamics for times
intermediate between two diverging time scales, $\ts \ll t \ll \ts'$.
It is important because of the connection it establishes between the
late part of $\beta$ relaxation to the early part of the
$\alpha$ relaxation. The latter deals with the plateau-below dynamics
on scale $\ts'$ and is globally governed by the {\em second scaling
  law} (also known as the time-temperature superposition principle):
\beq \Phi( t) = \tilde \Phi\left(\frac{t}{\ts'}\right).  \eeq
Here $\tilde \Phi$ is another $\sigma$-independent master function.
As an example, Figs.~\ref{fig.phi_raw}, \ref{fig.scaling1} and
\ref{fig.scaling2} show the typical two-step relaxation and the
associated \MCT\ scaling laws for the \CFS\ that we now describe.

{\bf \CFS}. -- In the dynamic facilitation approach the mesoscopic
structure of a liquid at temperature $T$ is represented by an assembly
of binary spins, $s_i=\pm 1$, $i=1,\,...,\,N$, whose value depends on
whether the local liquid density at site $i$ is higher or lower than
the average.  Energetic interactions are absent but the spin dynamics
is facilitated (or constrained)~\cite{FrAn}: At each time step a
randomly chosen spin $s_i$ is flipped with probability $w(s_i \to
-s_i) = {\rm min} \left\{ 1, {\rm e}^{- s_i/T} \right\}$, provided
that the spin $s_i$ is surrounded by at least ${\sf f}$ nearby
(liquidlike) up spins. This constraint mimics the cage effect: when
the temperature is low enough the fraction of liquidlike spins is
vanishingly small and, consequently, spin relaxation may involve a
large number of cooperative spin flips over regions with increasing
size.  Here, the relevant range of the facilitation parameter, ${\sf
  f}$, is $1 < {\sf f} < z-1$, where $z$ is the lattice coordination.
The correlator of our interest shall be the persistence $\Phi(t)$,
i.e., the probability that a spin has never flipped between times $0$
and $t$. Its long-time limit, the probability that a spin is
permanently frozen, is precisely the analog of the nonergodicity
parameter $f$. On the Bethe lattice this cooperative dynamics exhibits
features qualitatively similar to
\MCT~\cite{SeBiTo,ArSe,Se2012,FrSe,IkMi}, though one would, naively,
expect that the conventional form of \MCT\ fails for this type of model
as static correlations vanish. In fact, previous works on related
facilitated systems on finite dimensional lattices showed that
\MCT-like approximations are unable to describe the overall glassy
dynamics~\cite{Kawasaki,Pitts,Einax}.  This can be generally
understood through the connection of \CFS\ with the bootstrap
percolation transition which is smeared out in finite dimensions.  For
this reason, our evaluation of the \MCT\ status is carried out on the
Bethe lattice, which is the first natural step of a statistical
mechanics treatment. This is also relevant because: (i) \MCT\ scaling
regimes are sometimes hardly observed in numerical simulations of
disordered systems~\cite{Brangian,Sarlat}, and (ii) \MCT\ shows a
behavior which, in the limit of large space dimensionality, does not
conform to the replica theory~\cite{Ikeda,Schilling,Patrick}.

\begin{figure}
\begin{center}
\includegraphics{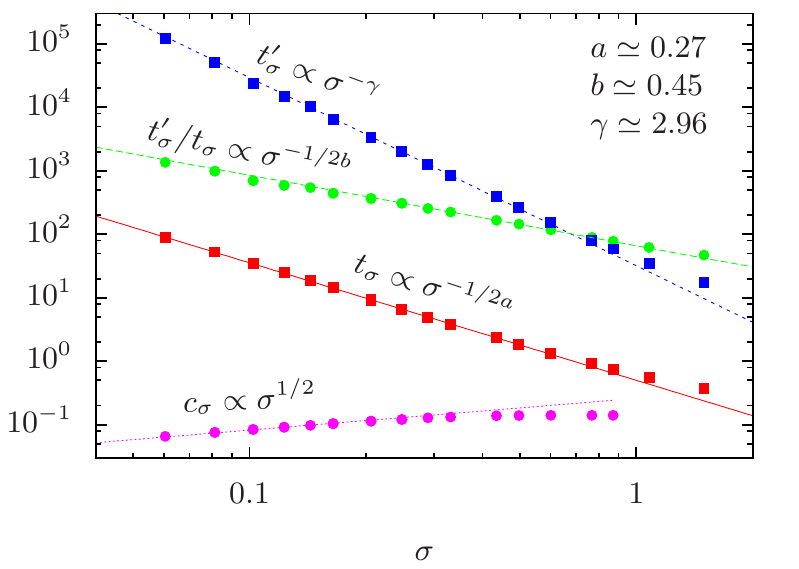}
\end{center}
\vspace{-.5cm}
\caption{Time scales $\ts$ and $\ts'$ and their ratio vs
  $\sigma=T/\Tc-1$ in the \CFS. Data for the correlation scale $\cs$
  are calculated exactly by using Eq.~(\ref{eq_cs_lambda}) with $\lambda
  \simeq0.815$. Straight lines are \MCT\ predictions.}
\label{fig.cs}
\end{figure}

To compare \CFS\ with \MCT\ we have first checked the $\beta$- and
$\alpha$-scaling laws along with Eqs.~(\ref{eq_cs}) and
(\ref{gamma_ab}).  We proceed as follows.  The time scale $\ts$ is
estimated as
\beq \ts = \int_0^{t^*} \left[ \Phi(t) - \Phi(t^*) \right]
\ \mathrm{d}t, 
\label{def_ts}
\eeq
where $t^*$ is the time to cross the plateau, $\Phi(t^*)=\fc$. This is
easily done as $\fc$ is known exactly. For the Bethe lattice we
consider here, $z=4$ and ${\sf f}=2$, we have $\Tc = 0.48089...$ and
$\fc = 0.67309...$\cite{SeBiTo}.
Figure~\ref{fig.cs} shows that $\ts$ obeys the power law $\ts \propto
\s^{-\frac{1}{2a}}$ with an exponent $a \simeq 0.27$. This value is
obtained by a fit over the temperature range $[0.52,\, 0.82]$.
Notice, that $a$ is the only critical exponent we estimate numerically
here. The correlation scale $\cs$, instead, is obtained from the
\MCT\ relation~\cite{Goetze_book}: 
\beq \cs = \sqrt{1-\lambda} \, (f - \fc) .  \label{eq_cs_lambda} \eeq
Here the jump of the order parameter, $f-\fc$, is deduced
from the exact calculation on the Bethe lattice~\cite{SeBiTo}, while
$\lambda$ is estimated by exploiting Eq.~(\ref{lambda_a}) which gives
$\lambda \simeq 0.815$. The result for the correlation scale $\cs$ is
shown in Fig.~\ref{fig.cs} along with the \MCT\ prediction.  The
excellent collapse of the rescaled relaxation data showed in
Fig.~\ref{fig.scaling1} is obtained by using the $\cs$ calculated in
this way.
We then get $b \simeq 0.45$ and $\gamma \simeq 2.96$ from
Eqs.~(\ref{lambda_b}) and (\ref{gamma_ab}). This latter value is
consistent with that found in Ref.~\cite{SeBiTo} ($\gamma \simeq 2.9$).
As a consistency check we then estimate the time scale $\ts'$ through
Eq.~(\ref{def_ts}) with $t^*$ such that
$\Phi(t^*)=0$. Figure~\ref{fig.cs} shows that $\ts'$ and the ratio
$\ts'/\ts$ are correctly predicted by \MCT.  From $\ts'$ we finally
get the second scaling law for the $\alpha$-relaxation process that we
find to hold in a wide range of temperatures above $\Tc$, see
Fig.~\ref{fig.scaling2}.

\begin{figure*}
\begin{center}
\includegraphics{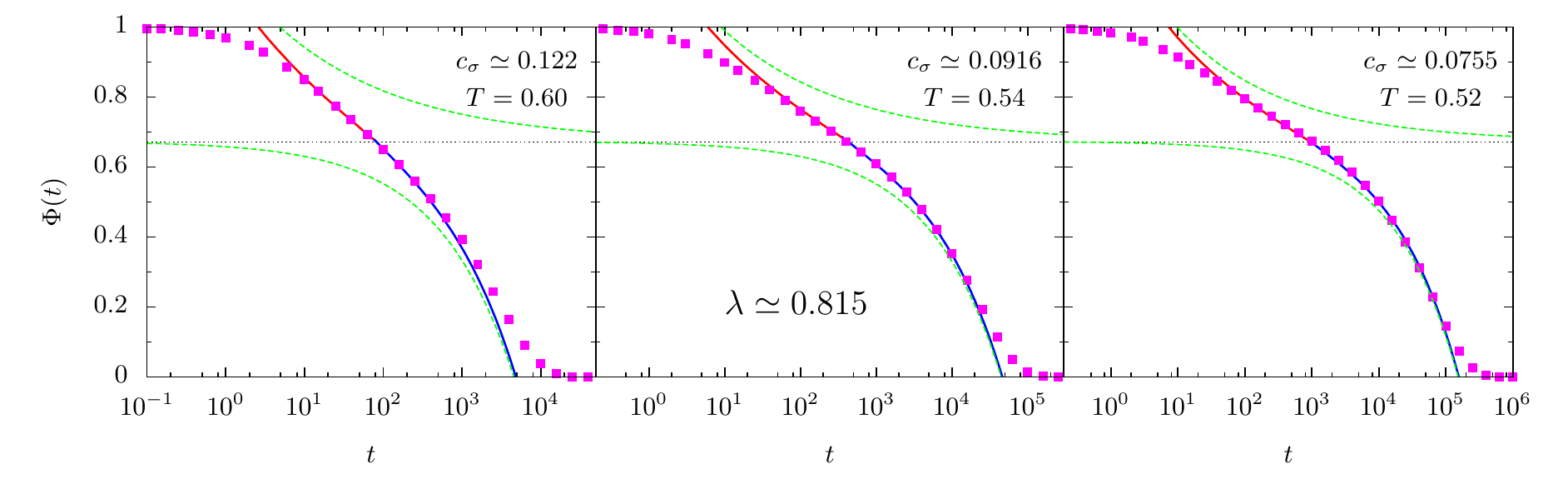}
\end{center}
\vspace{-.5cm}
\caption{Persistence $\Phi(t)$ vs time $t$ at temperature $T$ in the
  \CFS.  Full lines are the asymptotic leading corrections to the
  critical scaling laws predicted by \MCT.  The line above the plateau
  at $\fc = 0.67309$ represents the function $\fc + \cs \,
  g_a(\hat{t}/\hat{t}^*)$ while the blue line below the plateau
  corresponds to $\fc + \cs \, g_b(\hat{t}/\hat{t}^*)$.  For
  comparison, pure critical decay laws
are shown as dashed lines.}
\label{fig.phi_3}
\end{figure*}

We now consider in detail the \MCT\ predictions for the asymptotic
leading order corrections to the critical decay laws,
Eq.~(\ref{eq_gab}).  The coefficients entering Eq.~(\ref{eq_gab}) as
obtained from the critical exponents estimated above are $A_1 \simeq
1.58$, $B_1 \simeq 0.812$ and $B \simeq 1.32$.  We shall use these
values to test in a self-consistent manner the \MCT\ predictions: any
observed deviation would mean that \MCT\ does not hold for
\CFS\ either because the $\lambda$ estimated for our system is not
connected to the critical decay exponents $a$ and $b$ by
Eqs.~(\ref{lambda_a}) and~(\ref{lambda_b}), or because
Eqs.~(\ref{eq_gab}), (\ref{eq_A1}) and (\ref{eq_B1}) just do not apply
to \CFS.  We have explored a relatively wide range of temperatures
above $\Tc$.  Figure~\ref{fig.phi_3} shows the behavior of $\Phi(t)$
for some values of $T$ along with the \MCT\ predictions for the
crossover through the plateau.  The full line above $\fc$ represents
the function $ \fc + \cs \, g_a(\hat{t}/\hat{t}^*)$ while the
plateau-below function corresponds to $\fc + \cs \,
g_b(\hat{t}/\hat{t}^*)$. To appreciate the effect of leading order
corrections we also add the critical decay laws (which are obtained by
setting $A_1=B_1=0$) as dashed lines.  We see that, in agreement with
the nature of \MCT\ corrections, the quality of comparison increases
as $T$ decreases. In particular, up to the $80\%$ of the correlator
shape is accurately reproduced, over a time window ranging from two to
more than three decades; see Fig.~\ref{fig.phi_3}.
No significant improvement is achieved with the exact numerical
solution of Eq.~(\ref{eq_g})~\cite{nota}.
It is interesting to observe that, even though \MCT\ was not obviously
devised for describing \CFS, the \MCT\ predictions perform even better
than a recent truncation scheme for facilitated master equations; see
Fig.~4 in Ref.~\cite{Fennell} for a comparison.  Thus, it seems that
\MCT\ fully captures in a very general way the slow dynamics of large
scale cooperative rearrangements occuring near the glass singularity.

What makes the observed agreement pretty remarkable, however, is the
absence of any fitting parameter in our procedure.  In fact, one
should keep in mind that the prefactors of leading order corrections,
$A_1$ and $B_1/B$, are correlated and have no temperature dependence,
i.e., they are strongly constrained. Thus, to check the stability of
our comparison, we changed the exponent $a$ by an amount of $\pm
0.01$, that is essentially the uncertainty we have on this exponent.
We then found that relaxation behavior is still well reproduced if all
other quantities entering Eq.~(\ref{eq_gab}) are changed according to
\MCT\ relations. This robustness seems to suggest that prefactors of
leading order corrections are rather optimal from the purely data
fitting point of view.
Nevertheless, we notice that very close to $\Tc$ (below
$T=0.51$), the theoretical value of $\cs$ need to be increased by a
small factor (up to $5\%$) in order to accurately describe the
intermediate stage of $\alpha$ relaxation.
This discrepancy is presumably due to critical finite-size corrections
which are hardly accounted for in a \MCT\ description. The exact
solution of a related cooperative facilitated system should help to
settle this issue~\cite{PaSe}. Since static correlations are absent
here, we expect that a suitable \MCT\ formulation of facilitation
dynamics is related to those studied in Refs.~\cite{KaKi,ScSz}.

\medskip

{\bf Conclusions}. -- We have shown that the relaxation behavior of
cooperative facilitated spin models is accurately described, in a
relatively wide range of temperatures, by the \MCT\ predictions for
the asymptotic leading corrections to critical decay laws.  These
results support the idea of glassy universality and put on a solid
quantitative ground the correspondence between \MCT\ and \CFS.  At the
same time new challenges arise.

It is well known that the dynamics of disordered p-spin-like systems
is also exactly described by \MCT\, which is the key to the celebrated
random-first order transition (\RFOT) scenario for the glass
transition~\cite{RFOT_1,RFOT_2}.  \RFOT\ is based on a nontrivial
phase structure characterized by a one-step replica symmetry breaking
and a possible fractal free-energy landscape lying deeper in the glass
phase~\cite{Fractal}.  In the \CFS, instead, thermodynamics plays no
role: the configuration space breaks into several components that
cannot be connected by a sequence of allowed spin flips.  The duality
between kinetic and static representations seems therefore
irreconcilable. This would imply, for example, that the dynamical
inverse problem of inferring the interactions among system units
starting solely from time correlation data is undecidable.  However,
in view of the double connection (between \MCT\ and \CFS\, on one
hand, and between \MCT\ and \RFOT\, on the other) one is led to
conclude that a map connecting \CFS\ and \RFOT\ must necessarily
exist~\cite{Foini}.  Extending the mapping below the Kauzmann
temperature would obviously require extra thermodynamic ingredients
which are arguably provided by suitable energetic interactions, as
shown in Ref.~\cite{Gradenigo}.
Thus, the existence of three {\em a priori} distinct frameworks that
give the same dynamical scaling laws, while certainly surprising,
should not be considered as a mere coincidence, but rather as a genuine
signature of the universality of glassy relaxation captured by these
frameworks.
One advantage of the \CFS\ is that the glass formation can be
interpreted in real space as the formation of a bootstrap percolation
backbone of permanently frozen spins. Much less clear is how to
extract the values of the critical exponents $a$ and $b$ from this
geometrical structure, since no spin is permanently frozen on the
liquid side of the glass transition. So, at the moment this important
issue remains elusive in this framework as well as in other ones.
Another perspective suggested by the \CFS\ is that the
ergodicity-restoring activated events in finite dimensions should be
interpreted as the analog of the finite-volume metastability effects
which are known to transform the bootstrap percolation transition in a
(sharp) crossover~\cite{Aizenman}. Finally, it would be crucial to
extend the above investigation to finite dimensional facilitated
systems having a glass transition (see, e.g., Ref.~\cite{Sasa}).
The exploration of these problems is left to future works.

\newpage

W. G\"otze is gratefully acknowledged for elucidations about \MCT\ and
for his interest in this work. I also thank S. Franz and R. Schilling
for valuable comments on the manuscript.


\begin{thebibliography}{50}

\bibitem{BiKo} K. Binder and W. Kob, {\it Glassy Materials and
  Disordered Solids} (World Scientific, Singapore 2011).

\bibitem{Goetze_book} W. G\"otze, \emph{Complex Dynamics of
  Glass-Forming Liquids} (Oxford University Press, Oxford, 2009).

\bibitem{Pusey} W. van Megen, S.M. Underwood and P.N. Pusey,
  Phys. Rev. Lett. {\bf 67}, 1586 (1991).

\bibitem{KoAn} W. Kob and H.C. Andersen, Phys. Rev. Lett. {\bf 73},
  1376 (1994).

\bibitem{Goetze90} W. G\"otze, J. Phys. Condens. Matter {\bf 2},
  8485 (1990)

\bibitem{Franosch} T. Franosch, M. Fuchs, W. G\"otze, M.R. Mayr,
  and A.P. Singh, Phys. Rev. E {\bf 55}, 7153 (1997).

\bibitem{Gleim} T. Gleim and W. Kob, Eur. Phys. J. B {\bf 13}, 83
  (2000).

\bibitem{FrAn} G.H. Fredrickson and H.C. Andersen, Phys. Rev. Lett.
  {\bf 53}, 1244 (1984); J. Chem. Phys.  {\bf 83}, 5822 (1985).

\bibitem{SeBiTo} M. Sellitto, G. Biroli, and C. Toninelli,
  Europhys. Lett. {\bf 69}, 496 (2005).

\bibitem{ArSe} M. Sellitto, D. De Martino, F. Caccioli, and
  J.J. Arenzon, Phys. Rev. Lett. \textbf{105}, 265704
  (2010); J.J. Arenzon and M. Sellitto, J. Chem. Phys. \textbf{137},
  084501 (2012).

\bibitem{Se2012} M. Sellitto, Phys. Rev. E \textbf{86}, 030502(R)
  (2012); J. Chem. Phys. {\bf 138}, 224507 (2013).

\bibitem{FrSe} S. Franz and M. Sellitto, J. Stat. Mech. (2013) P02025.

\bibitem{IkMi} H. Ikeda and K. Miyazaki, Europhys. Lett. {\bf 112}
  16001 (2015).

\bibitem{Kawasaki} K. Kawasaki, Physica A (Amsterdam) {\bf 215}, 61
  (1995).

\bibitem{Pitts} S.J. Pitts, T. Young, and H.C. Andersen,
  J. Chem. Phys. {\bf 113}, 8671 (2000).

\bibitem{Einax} M. Einax and M. Schulz, J. Chem. Phys. {\bf 115}, 2282
  (2001).

\bibitem{Ikeda} A. Ikeda and K. Miyazaki, Phys. Rev. Lett. {\bf 104},
  255704 (2010); {\bf 106}, 049602 (2011).

\bibitem{Schilling} B. Schmid and R. Schilling, Phys. Rev. E {\bf 81},
  041502 (2010); R. Schilling and B. Schmid, Phys. Rev. Lett.  {\bf
    106}, 049601 (2011).

\bibitem{Patrick} P. Charbonneau, A. Ikeda, G. Parisi, and F. Zamponi,
  Phys. Rev. Lett. {\bf 107}, 185702 (2011)

\bibitem{Brangian} C. Brangian, W. Kob, and K. Binder, J. Phys. A {\bf
  35}, 191 (2002).

\bibitem{Sarlat} T. Sarlat, A. Billoire, G. Biroli, and
  J.-P. Bouchaud, J. Stat. Mech. P08014 (2009).

\bibitem{nota} The numerical code for solving Eq.~(\ref{eq_g}) has
  been kindly provided by Th. Voigtmann.

\bibitem{Fennell} P.G. Fennell, J.P. Gleeson, and D. Cellai, Phys. Rev. E
  \textbf{90}, 032824 (2014)

\bibitem{PaSe} G. Parisi and M. Sellitto, Europhys. Lett. {\bf 109},
  36001 (2015).

\bibitem{KaKi} K. Kawasaki and B. Kim, Phys. Rev. Lett. {\bf 86}, 3582
  (2001); J. Phys. Condens. Matter {\bf 14}, 2265 (2002).

\bibitem{ScSz} R.Schilling and G. Szamel, Europhys. Lett. {\bf 61},
  207 (2003); J. Phys. Condens. Matter {\bf 15}, S967 (2003).

\bibitem{RFOT_1} V. Lubchenko and P.G. Wolynes,
  Annu. Rev. Phys. Chem. {\bf 58}, 235 (2007).

\bibitem{RFOT_2} T.R. Kirkpatrick and D. Thirumalai,
  Rev. Mod. Phys. {\bf 87}, 183 (2015).

\bibitem{Fractal} P. Charbonneau, J. Kurchan, G. Parisi, P. Urbani,
  and F. Zamponi, Nat. Commun. {\bf 5}, 3725 (2014).

\bibitem{Foini} For a relatively simpler case in which this mapping
  can be effectively realized, see L. Foini, F. Krzakala, and
  F. Zamponi, J. Stat. Mech. (2012) P06013.

\bibitem{Gradenigo} S. Franz, G. Gradenigo, and S. Spigler,
  arXiv:1507.05072

\bibitem{Aizenman} M. Aizenman and J. L. Lebowitz, J. Phys. A {\bf
  21}, 3801 (1988).

\bibitem{Sasa} H. Ohta and S. Sasa, J. Stat. Phys. {\bf 155}, 827
  (2014).

\end{thebibliography}
\end{document}